\def\msol{{\rm\thinspace M${}_\odot$}}
\def\fig{{\rm\thinspace Figure}}

\def\etal{{\it et al.\thinspace}}
\def\eg{{\it e.g.\ }}

\def\ie{{\it i.e.\ }}

\def\msolyr{{\rm\thinspace M${}_\odot\; {\rm yr}^{-1}$}}
\def\gcm3{{g cm${}^{-3}$}}
\def\km{{\rm\thinspace km}}

\def\msol{\hbox{$\rm\thinspace M_{\odot}$}}
\def\msolyr{\hbox{$\rm\thinspace M_{\odot} \; {\rm yr}^{-1}$}}

\def\s{{\rm\thinspace s}}

\def\h50{\hbox{$\rm\thinspace h_{50}$}}
\def\h50m1{\hbox{$\rm\thinspace h_{50}^{-1}$}}
\def\kmps{\hbox{$\km\s^{-1}\,$}}

\def\s{\mathop{\rm s}\nolimits}

\def\km{\mathop{\rm km}\nolimits}

\def\msol{\hbox{$\rm\thinspace M_{\odot}$}}
\def\etal{{\it et al.\thinspace}}
\def\eg{{\it e.g.\ }}

\def\ie{{\it i.e.\ }}

\def\fig{figure}

\def\p3m{P${}^3$M}
\def\ap3m{AP${}^3$M}
\def\-{{\em{---}}}

\documentclass{article}
\usepackage{emulateapj}

\begin{document}
\title{High Resolution Simulation of Galaxy Formation with Feedback}
\author{R. J. Thacker\altaffilmark{1,2,3} and H. M. P.
Couchman\altaffilmark{2,4}}
\authoremail{thacker@astro.berkeley.edu}

\altaffiltext{1}{Theoretical Physics Institute, Department of Physics,
University of Alberta, Edmonton, Alberta, T6G 2J1, Canada}
\altaffiltext{2}{Department of Physics and Astronomy, 
University of Western Ontario, London, Ontario, N6A 3K7, Canada.}
\altaffiltext{3}{Current address: Department of Astronomy, 
University of California at Berkeley,
Berkeley, CA, 94720.}
\altaffiltext{4}{Current address: Department of Physics and Astronomy,
McMaster University, 1280 Main St. West, Hamilton, Ontario, L8S 4M1,
Canada.}

\begin{abstract}
We present results from a Smoothed Particle Hydrodynamic (SPH) simulation
of galaxy formation that exceeds the minimum resolution requirement
suggested by Steinmetz \& Muller (1993) of $3\times10^4$ SPH particles per
galaxy.  Using the multiple mass technique an effective resolution of a
little over one billion particles is attained within a 48 Mpc cube.  We
find that even with an SPH mass resolution of $1.5\times10^6$ \msol\ and a
plausible feedback algorithm, the cooling catastrophe continues to be a
problem for Einstein-de Sitter CDM cosmologies. Increasing resolution also
appears to exacerbate the core-halo angular momentum transport problem. 

\end{abstract}

\section{Introduction} 
Steinmetz \& Muller (1993, hereafter SM93) have shown that in SPH
simulations at least $3\times10^4$ SPH particles are necessary to
accurately follow local hydrodynamic evolution, in particular shocks and  
velocity fields. The simulation we discuss in this {\em Letter} is the
first simulation of galaxy formation, we are aware of, to meet this
criterion and also account for long range tidal fields. The tidal fields
are incorporated by using the multiple mass technique
(Porter 1985). 

To date, the highest resolution studies of galaxy formation including long
range tidal forces are those of Navarro and Steinmetz (1997) which used up
to 5,000 gas particles per galaxy. These simulations were integrated to
$z=0$ and required over 50,000 time-steps. Although the particle number is
comparatively low, the integration to $z=0$ is
a significant achievement. 

The simulation presented here includes star formation and feedback using
an algorithm tested in detail in Thacker \& Couchman (1999, hereafter
TC99).  Many authors believe (\eg White 1994) that feedback may solve both
the cooling catastrophe (White \& Frenk 1991) and the associated problem
of core-halo angular momentum (AM) transport (Navarro \& Benz 1991). The
morphological results for objects in this simulation are of great interest
since we are able to probe smaller mass scales than previous
investigations. This is particularly relevant in simulations with feedback
since lower mass halos are more susceptible to perturbations produced by
feedback.

\section{Algorithm and Initial conditions}
The Bond \& Efstathiou (1984) CDM power spectrum was used to assign
curvature perturbations in an Einstein-de Sitter universe, with
cosmological parameters $\Omega_b=0.1$, $\Omega_{CDM}=0.9$, $h=0.5$, shape
parameter $\Gamma=0.41$ and normalization $\sigma_8=0.6$. The
perturbations were assigned to a low resolution $100^3$ dark-matter-only
simulation of comoving width 48 Mpc, at a redshift of $z=67$. The
simulation was evolved to $z=1$ using the adaptive P${}^3$M algorithm of
Couchman (1991), at which point a halo of mass $1.66\times 10^{12}$
\msol\, was selected for re-simulation. The halo, which lies on a filament
approximately 2 Mpc long, does not have a violent merger history at this
resolution and can be categorized as a being the halo of a field galaxy.

The initial conditions of the high resolution simulation were prepared by
creating four mass hierarchies in radial shells within the simulation
volume. Each shell has a width half that of the previous hierarchy, and
the per-particle mass scales by a factor of 8 between each level, yielding
a central high resolution region 6 Mpc in comoving diameter. The mass in
this region is $7.8\times10^{12}$ \msol, and a particle number of
$2\times$523,535 gives particle masses of $1.5\times 10^6$ \msol\, and
$1.4\times 10^7$ \msol\, for gas and dark matter respectively. The minimum
`glob' and dark halo mass resolutions are 52 times higher, corresponding
to the number of neighbors in the SPH solver. Only the high resolution
region includes SPH particles, which were given an initial temperature of
1,000 K. The effective resolution of the high resolution region is
$2\times 800^3$.  The particle positions were drawn from a `glass' and the
same power spectrum was used to assign modes to the particle distribution
although the spectrum was truncated at the Nyquist frequency of each
hierarchy. In the high resolution simulation at $z=1$ the candidate halo
is represented by about 220,000 particles, half dark matter and half gas.
A Plummer softening length of $\epsilon=1.5$ kpc was chosen and the
minimum SPH smoothing length was $h_{min}=1.76$ kpc. 

In TC99 it was demonstrated that the temperature smoothing (TS) algorithm
produces the most significant feedback effect. However, it was also
observed that the NGC 6503 prototype was more affected by feedback than
was the Milky Way prototype.  Hence, given the higher resolution in this
simulation, the energy smoothing (ESa)  algorithm was adopted. The ESa
algorithm smooths $5\times 10^{15}$ erg g${}^{-1}$ of feedback energy over
the neighbor particles of an SPH particle after a star formation event and
allows this energy to persist for a time $t^*=5$ Myr.  Given the higher
densities resolved in the simulation the Schmidt Law SFR normalization was
reduced by 30\% compared to the low resolution runs in TC99. A
self-gravity criterion, also prevents star formation in regions where
$\rho_b<0.4\,\rho_{DM}$.

\section{Results}
It was not possible to integrate the simulation beyond $z=2.16$ due to
inefficiencies developed in the SPH algorithm as a result of the fixed
minimum smoothing length employed.  Integration to this redshift required
4,100 time-steps; approximately 37,000 would be required to $z=0$. By
$z=2.16$ 40,777 star particles had been created (4\% of the initial gas
mass) and $r_{200}=75$ kpc. 

The structure and evolution of the gas in the simulation is depicted in
\fig~4. Three notable stages of evolution are evident: (i)  by $z=10$,
five gas cores have formed with masses between $10^8$ \msol\ and $6\times
10^8$ \msol\ and the associated dark halo masses are between $10^9$ \msol\
and $5\times 10^9$ \msol. A small fraction of the gas has been shocked to
temperatures greater than 5,000 K, but otherwise adiabatic cooling due to
expansion dominates; (ii) by $z=5$, well over 100 dark matter halos and
globs are resolved. The largest dark matter halo at this epoch has a mass
of $5\times10^{10}$ \msol\ and the associated glob $5\times10^{9}$ \msol. 
Between $z=3$ and $z=5$, a large fraction of halos merge and it becomes
clear that accretion on the central object is dominated by collapse along
a filament in the z-direction.  The hot gas halo begins developing at this
epoch;  (iii)  by $z=2.16$, the hot halo has evolved significantly and has
a central temperature close to $10^6$ K (see section~\ref{halo}). The
largest dark matter halo has a mass of $6\times10^{11}$ \msol, while the
largest glob has a mass of $6\times10^{10}$ \msol. 

\subsection{Morphology and the effect of feedback} 
In our low resolution studies (TC99) disks form a very dense gas core,
with a spatial extent smaller than $0.1h_{min}$, in nearly all
simulations, the exception being the highly energetic temperature
smoothing feedback. There is a clear trend toward higher specific angular
momenta in the gas compared to dark matter with increased feedback.
However, the effect is small when compared to that needed to produce
observed spiral galaxy characteristics (Fall 1983). Increasing feedback,
in an attempt to unbind the dense gas cores, lead to hot halo gas being
unable to cool within a Hubble time, thus rendering it unavailable for
disk formation.

The higher resolution in this simulation, in combination with hierarchical
clustering, leads to the the gas overcoming the self-gravity criterion,
and hence forming stars, at an earlier epoch ($z=5$) than in the
simulations of TC99 ($z=4$). Note that the 1.5 kpc resolution allows for
density values that are 20 times higher than the simulations in TC99. 

The dwarf systems formed here did not have the tight central gas core,
observed in the lower resolution runs (TC99). The two largest dwarf
systems contained over 30,000 and 20,000 gas particles respectively,
before merging together at $z=2.16$. Visualization of the dwarfs shows
that feedback causes small pockets of hot gas which remain static until
the region reaches the end of the feedback period at which point the gas
cools rapidly. There is little evidence for `blow-out'.

In \fig~1, the distribution of dark matter, gas and stars at $z=2.16$ is
shown. The dark matter exhibits overmerging, which should be expected,
given that the gravitational force is only resolved to $0.02\,r_{200}$. 
The dark matter is comparatively featureless while the gas shows a number
of dense cores.  The results of Moore \etal (1998)  suggest that to avoid
overmerging, a force resolution of $0.002\,r_{200}$ is necessary, \ie ten
times smaller than that used here. 

The high mass resolution in the simulation allows us to resolve tidal
tails extremely well.  The viscosity of gas tends to accentuate the
stripping effect, with collisionless matter being much less susceptible. 
An examination of the final state, shown in \fig~1, shows that the bulk of
the gas objects merging with the central core are tidally stripped as they
merge. The largest dwarf system develops a ring feature because an
in-falling satellite becomes phase-wrapped as it accretes. 
{\epsscale{0.95}
\plotone{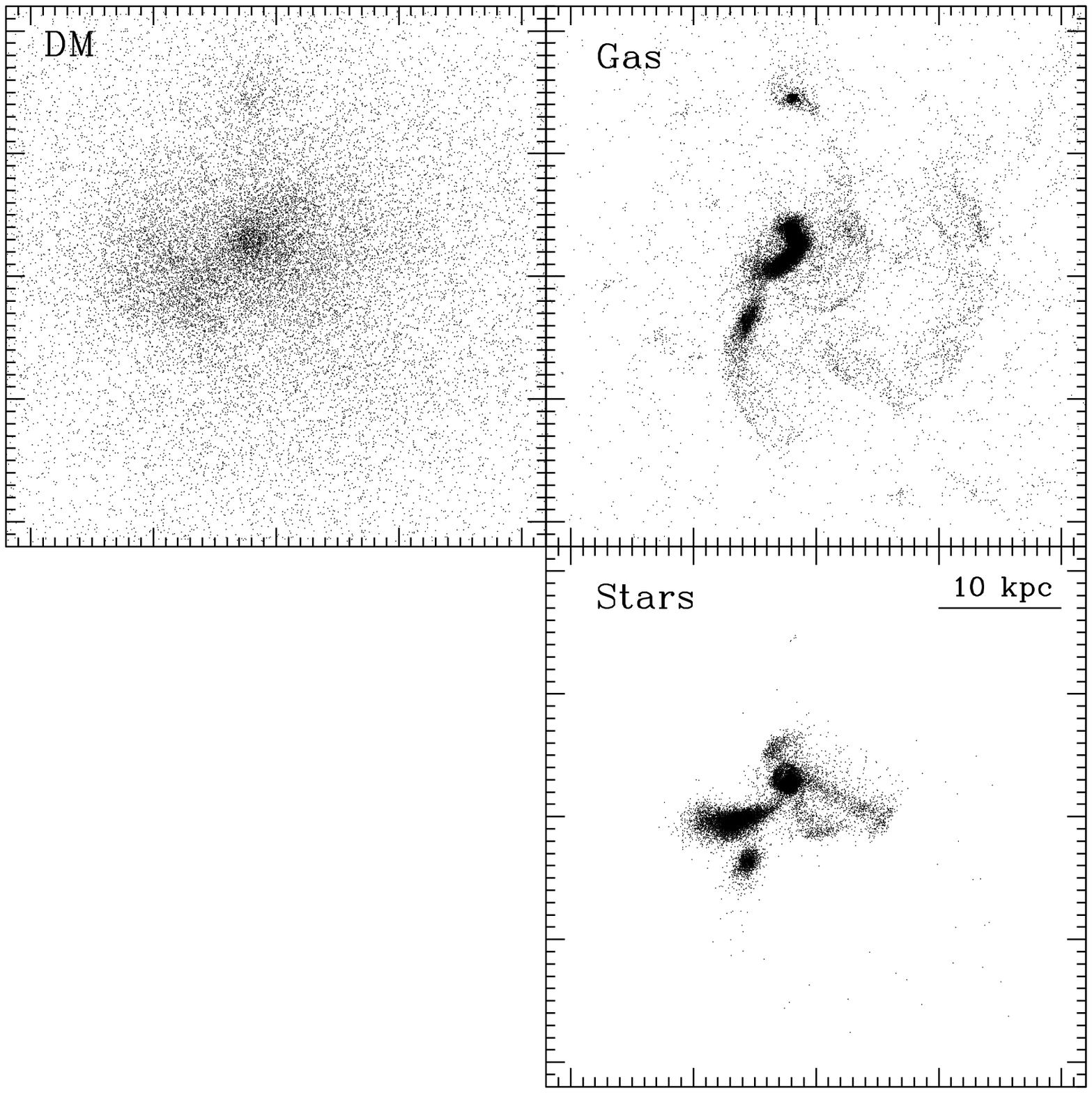}}

{\small {\sc Fig.}~1.---Z-projection of the dark
matter, gas and stellar distributions in the main halo at $z=2.16$.
Overmerging is apparent in the dark matter and it is not possible to
associate halos directly with the stellar and gaseous features.}  
\bigskip

\subsection{Angular momenta of the dwarfs and main halo} 

To analyze the growth of the specific angular momentum, $\bf{L}$, in the
two largest dwarf systems, the z-component of $\bf{L}$ was compared to the
expected value of a flattened rotating disk of radius $R$ and circular
velocity $v_c$. For a disk system with radial orbits $L_z=Rv_c$. The
analysis was performed at $z=2.2$ since the dwarfs later coalesce. For
both of the systems, most of the gas and star particles have $L_z$ values
marginally under the $Rv_c$ prediction, but do seem to follow the shape of
the predicted curve reasonably well. This suggests that both of the disks
have not yet lost a significant amount of angular momentum due to bar
formation, although this is likely to occur.  Note that the two dwarfs
have well-defined disks, albeit with a comparatively low aspect ratio
since the disk thickness is about 1 kpc, while the diameter is about 4 kpc
(which is smaller than $4h_{min}$). Analysis of the $X_2(R)$ (\cite{T81})
data show that both disks achieve stability at around a radius of 2 kpc,
which must be considered sub-resolution.

The same $L_z$ analysis was applied to the baryon condensation in the main
halo at $z=2.16$. Although no clear disk is yet visible, visualization
shows that a number of in-falling systems are orbiting in a similar plane.
Provided that these systems contribute the largest fraction of the orbital
component of $\bf L$ to the main halo, the dominant angular momentum
component should be perpendicular to this plane. The data show that almost
all the in-falling matter, picked out in the horizontal plane
perpendicular to $L_z$, has lost a significant proportion of angular
momentum relative to the $Rv_c$ prediction. Since no disk has formed, all
of this angular momentum loss must be due to the core-halo transport
mechanism. This result appears to show that at higher resolution the
angular momentum loss is greater. However, caution should be emphasized in
interpreting this result: if the selected matter just happens to be
passing through the plane then its $\bf L$ vector will not align well with
that of the entire system and hence the $L_z$ analysis will overestimate
the \\

{\epsscale{0.95}
\plotone{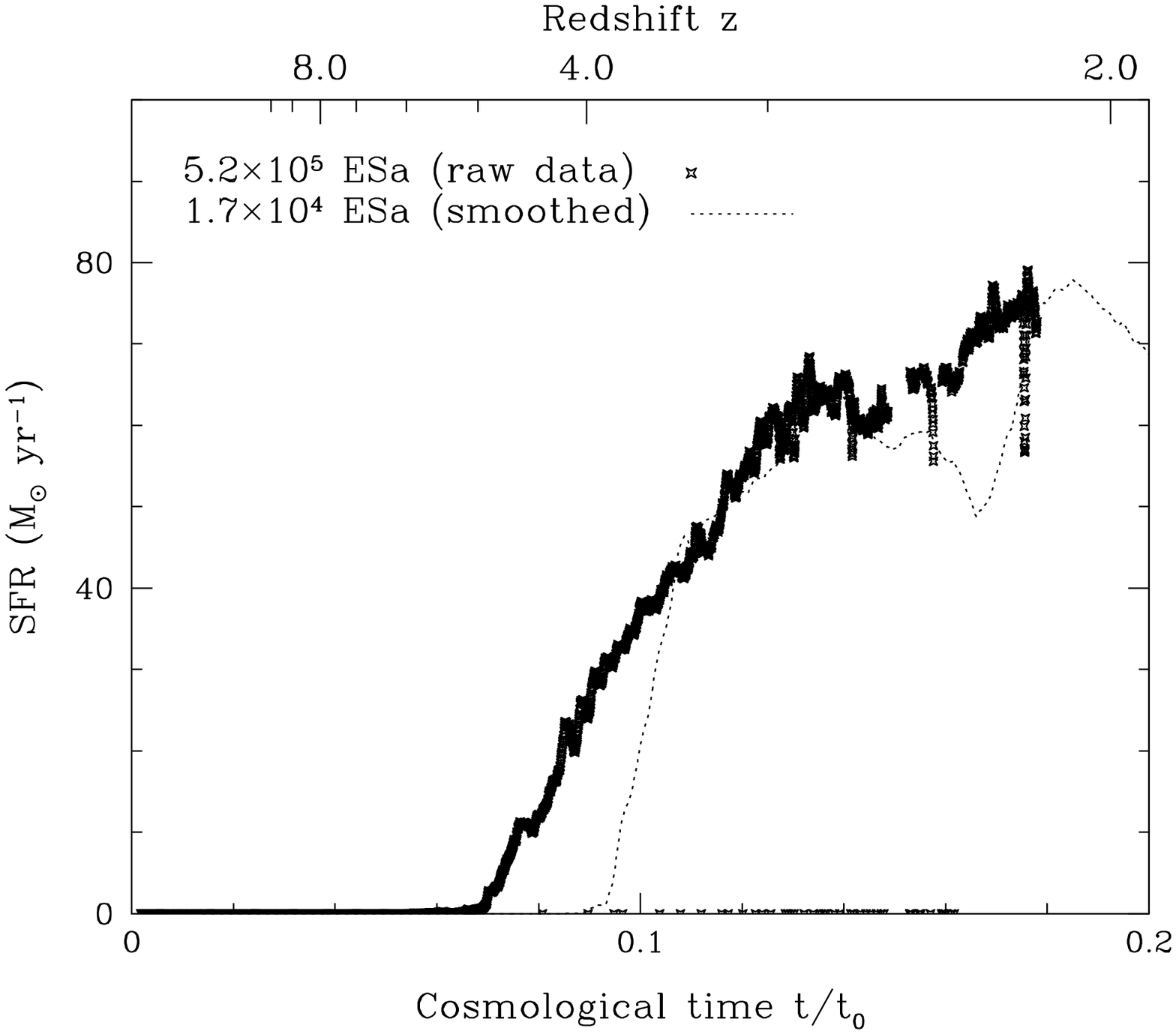}}

{\small {\sc Fig.}~2.---SFR integrated over the entire high resolution
region.
The higher mass resolution present in the simulation leads to a
smoother SFR than the low resolution data (the smoothed version
is shown for comparison). Points along the bottom of the plot are from   
code restarts and the small gap in the data is due to a log file being
accidentally erased.} \bigskip

\noindent loss of angular momentum. 
 
At $z=2.16$, the largest part of the specific angular momentum within
$r_{200}$ is carried by one system very close to $r_{200}$. Calculation of
$|{\bf L}|$ for the dark matter, gas cores and gas halo shows that all the
values lies within a factor of two. However, recalculating these values
within $r_{200}/2$, \ie removing the contribution from $r\sim r_{200}$,
shows markedly different results: the ratio of $|{\bf L}|$ of the stars to
dark matter is 0.19, compared to values in the range 0.1 to 0.15 for the
low resolution simulations at $z=1$ at $r_{200}$ in TC99. Hence the
angular momentum loss in the core region is occurring earlier in this high
resolution simulation. 

\subsection{Halo and glob mass multiplicity functions}
Three red-shifts were selected for analysis, namely $z=10,5,2.16$. To
identify halos the `friends of friends' group-finding algorithm was
employed.  The linking length for dark matter was $r_{DM}=0.15d$ while for
the baryons it was $r_{bary}=0.11d,0.06d,0.03d$ respectively for the
different red-shifts, where $d$ is the average inter-particle spacing. The
cumulative mass multiplicity functions for the dark matter and baryons are
fit well by $N(<M)\propto M^{-1}$. This is shallower than the observed
$M^{-2}$ power law in Evrard \etal (1994), but is in close agreement with
the results of Ghigna \etal (1999) who derive a (non-cumulative) mass
multiplicity function with power law slope $dN(M)/dM \propto M^{-1.9}$ in
a galaxy cluster simulation. Note, our results, and those of Ghigna \etal,
are measured close to a density peak and are thus biased. Further, the
tilt in the power law cannot be due to feedback blowing apart small baryon
cores since both the dark matter and the baryons exhibit a similar slope. 
There is noticeably more evolution in the globs between $z=5$ and $z=2.16$
than there is for the dark matter halos, which is probably related to the
Rees-Ostriker (1977) cooling criterion: prior to this epoch globs have not
been able to cool. 

\subsection{Star formation rate}\label{sfrdiscuss}
Although gas cores are beginning to form at $z=10$, none of them overcomes
the self-gravity criterion until $z=5$ at which point star formation
begins.  The higher resolution in this simulation leads to an integrated
SFR that is less burst-like than in the low resolution results of TC99.
The gradient of the SFR versus time is shallower than the lower resolution
runs since the SFR normalization is lower. As is shown in the plot of SFR
versus time in \fig~2, the peak SFR is reached at $z=2.18$ and is 80
\msolyr. A linear scaling of the peak value suggests an SFR of over 100
\msolyr\ would be attained if the low resolution SFR normalization from
TC99 had been kept.  The formation of the first star particles, and hence
the first feedback events, occurred at $z=3.4$ which is only slightly
earlier than the low resolution runs ($z=3.0$). This is due to the lower
SFR normalization.  Since there is no sudden drop in the SFR following the
first feedback events, energy smoothing does not have a significant effect
on the SFR at this resolution (as compared to the results found in the low
resolution temperature smoothing, and single particle feedback
experiments).

\subsection{Halo properties}\label{halo}
Although the center of the dark halo is not completely relaxed at $z=2.16$
(the dwarfs are merging) it is still interesting to plot the radial
density profile of the system since there is much interest in the shape of
halo profiles (\cite{NFW97,BM98}). 

The densities of the dark matter and gas are shown in \fig~3.  A fit of
the dark matter to the Moore \etal profile is shown for reference and the
fit is excellent.  The averaged SPH density (not shown) peaks about half
an order of magnitude higher than the dark matter density.  The
self-gravity criterion is achieved out to 4$\epsilon$, indicating that a
large fraction of the condensing gas is available for star formation. 

The radial temperature profile is approximately flat, \ie isothermal, with
a temperature of $6\times 10^5$ K out to a radius of 150 kpc (ignoring gas
for which $\delta_{gas}>2000$, which is assumed to be cold dense gas in
dwarf systems). The temperature declines steeply at a radius of 250 kpc,
where it falls from $2\times10^5$ K to $10^4$ K. The sound crossing time
is 0.03 Gyr so the hot gas distribution has had time to relax.  At the
center of the halo the cooling times are close to 10 Gyr, so cooling has
little effect on the profile.  The full three dimensional velocity
dispersion for the dark matter within $r_{200}$ was found to be 163 \kmps. 
The average temperature for all of the gas within $r_{200}$ was
$2.4\times10^5$ K, while for the hot halo it was $6.4\times10^5$ K. This
leads to isothermal $\beta$ parameters of $\beta_{all}$=1.89 and
$\beta_{halo}=0.73$. The value for the halo gas is lower than unity which
is consistent with energy input from feedback.

\section{Conclusion}

This {\em Letter} has presented results for a simulation with sufficient
mass resolution to accurately resolve shock structures within the forming
galaxy. Unfortunately, due to limitations in the simulation algorithm, it
was necessary to truncate the evolution of the system at $z=2.16$.

Principal conclusions follow:

1. The cooling catastrophe continues to be a significant problem. The
dwarf galaxies, even those containing greater than $10^4$ gas particles,
collapsed to a size close to the gravitational softening length of the
simulation. However, the dwarfs did not exhibit a very tight central
concentration of gas as observed in earlier low resolution models. Even
given the higher mass resolution in this simulation, ESa feedback still 
does not cause \\

{\epsscale{0.93}
\plotone{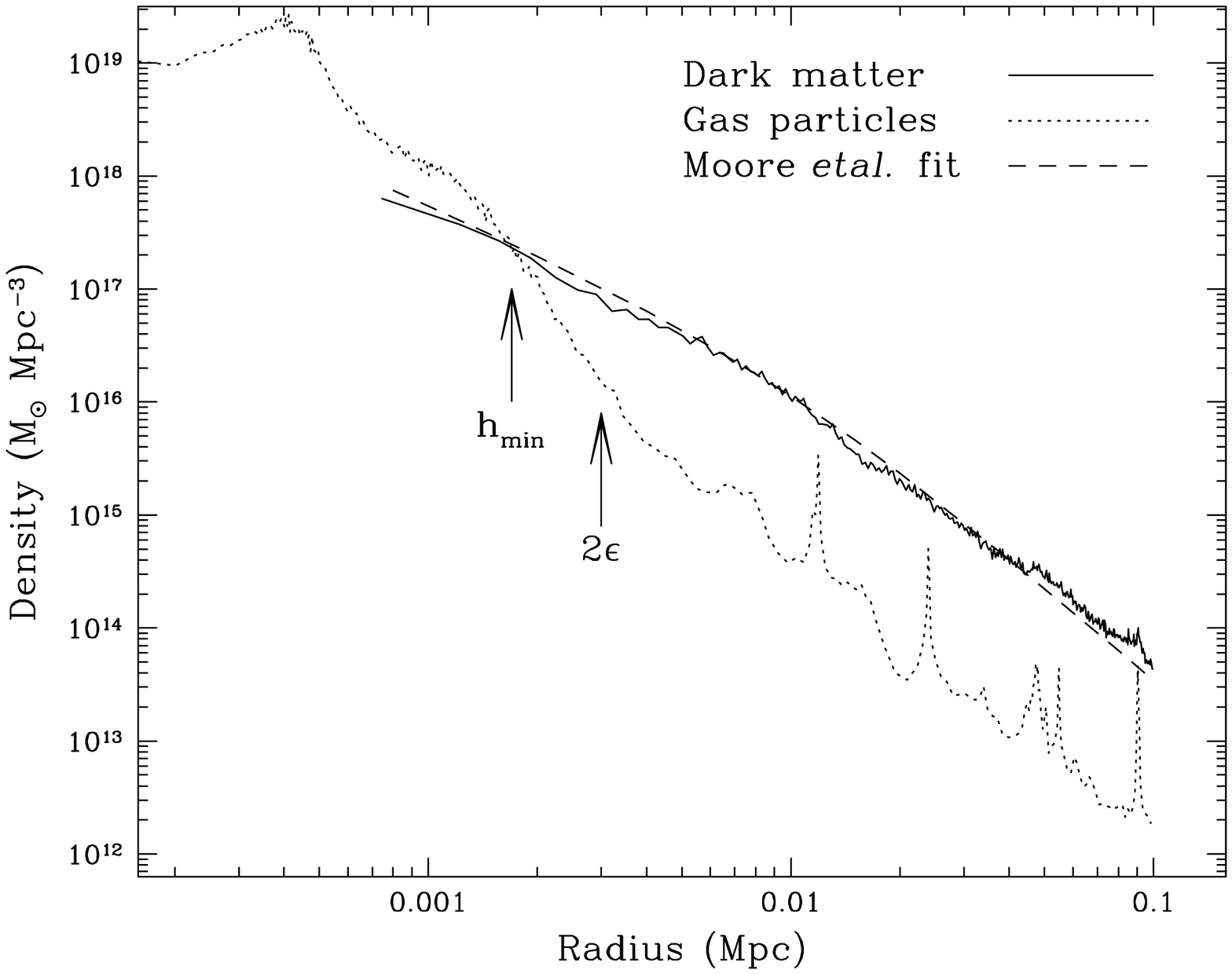}}

{\small {\sc Fig.}~3.---Density profiles for the dark matter and gas in
the main halo. The profiles are constructed using spherical 208 particle
Lagrangian bins.
}\bigskip

\noindent strong damping of the SFR. 

2. Overmerging is still observed. Including the baryons does not have
any significant effect on this problem. In the core of the main halo, a
number of the baryon cores do not have an accompanying dark matter halo.

3. The core-halo angular momentum transport mechanism remains a serious
problem. The star particles, which are formed from the
gas cores, showed a noticeable loss of specific angular momentum relative
to the dark matter at $z=2.16$.

4. The mass multiplicity function for both the dark matter halos and
globs is well fit by an $M^{-1}$ power law. Although it is tempting to suggest this
change is due to feedback reducing the number of low mass objects,
this not the case.

The higher resolution moved the onset of star formation to $z=5$, compared
to the $z\simeq 3.5$ epoch for the low resolution runs in TC99.
Nonetheless $z=5$ is still later than some of the observations suggest. 
For example, Chen \etal (1999) report the possible identification of a
star-forming galaxy at $z=6.68$. The estimated star formation rate is 70
\msolyr, assuming a flat, $h=0.5$ cosmology. Given the simulation just
presented, this result seems remarkable. Even though the self-gravity
criterion delays star formation until comparatively late times, it is
difficult to see how, with the power spectrum used, such an object could
be formed. The earlier onset of star formation can be achieved by adding
more resolution or by using a power spectrum with a higher normalization. 
However, continuing to add resolution has limits since 100 times the
current mass resolution would enable the Jeans' Mass of the first objects
in the CDM cosmology to be resolved, \ie there is a limit to the cooling
catastrophe.  We note that the Lyman break galaxies, characterized by
masses of a few $10^{10}$ \msol, and star formation rates of order 5-25
\msolyr, are quite well approximated within the simulation. The two large
dwarfs could conceivably be likened to these objects.

Changes are currently being made to our simulation algorithm and we hope
to evolve this simulation further with the new code. Nonetheless this
simulation provides one particularly important result: even at a mass
resolution of $1.5\times10^6$ \msol, a plausible feedback algorithm still
fails to prevent the cooling catastrophe. 

\acknowledgements
The authors thank Jimmy Scott of SGI-Cray Canada for securing a grant of
supercomputer time at the Eagan Supercomputing Center where part of this
research was conducted. A grant of time on the UK-CCC server, `COSMOS',
provided by the Virgo Consortium is also acknowledged.  RJT was supported
by a Dissertation Fellowship from the University of Alberta while this
research was conducted. HMPC thanks NSERC of Canada for financial support.

\newpage
{\epsscale{1.0}
\plotone{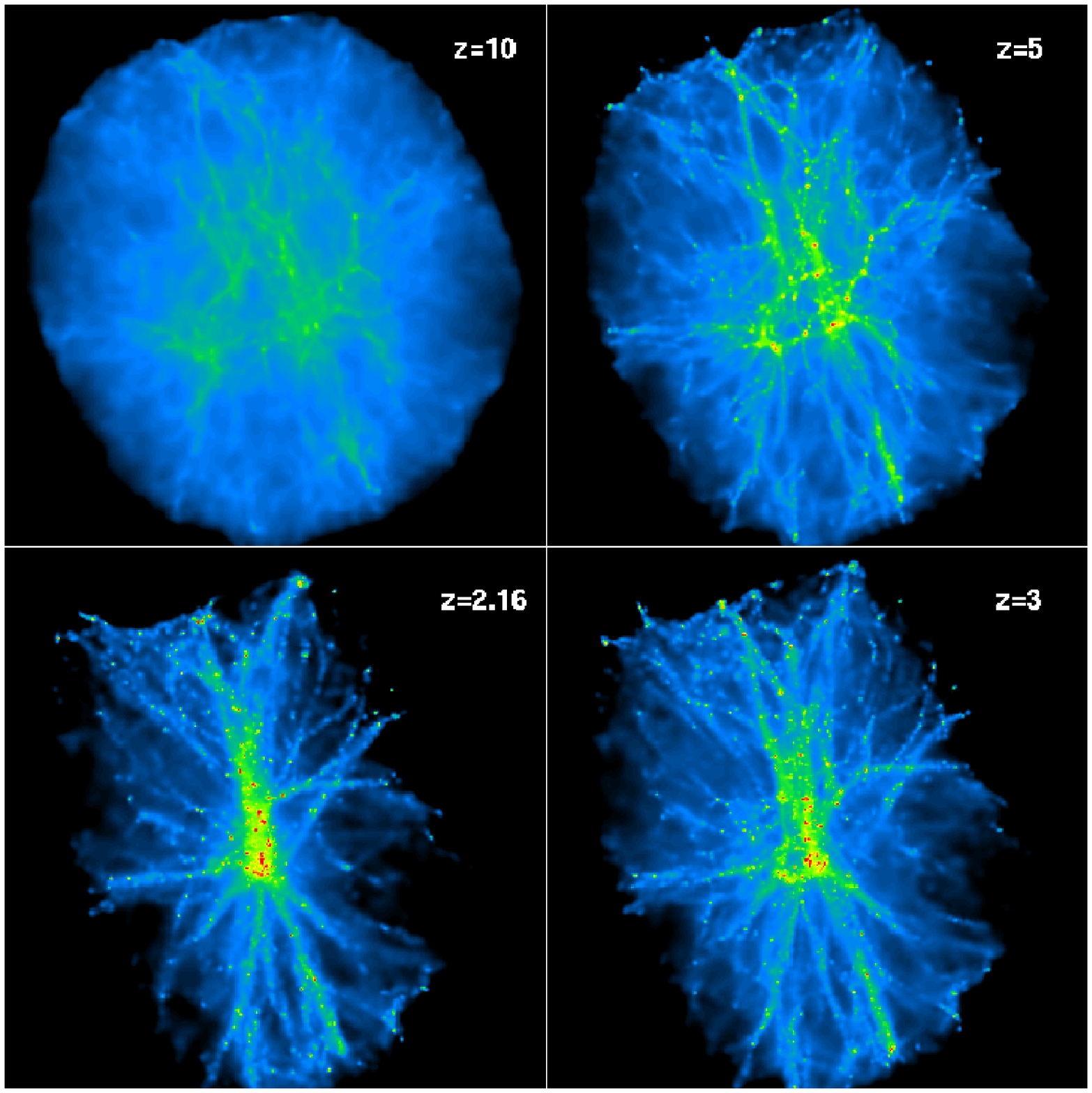}}

{\small {\sc Fig.}~4.---4-panel plot showing the evolution 
of the comoving density from $z=10$ to
the final epoch $z=2.16$. The SPH data is smoothed onto a grid with
spacing $h_{min}$, thus the grids are a realistic representation of the
resolution. In physical coordinates the top left panel would be 3.48
times
smaller than the bottom left.  The color scheme runs from $10^{18}\;
n_B\; {\rm cm}^{-2}$ (blue) to $10^{21}\; n_B \;{\rm cm}^{-2}$ (red).
The filamentary structure is already forming at $z=10$ and by $z=5$ the
first halos have reached sufficient density to form stars. Evolution
from
$z=3$ to $z=2.16$ is dominated by collapse along the x-direction. Note
that the z-projection looks directly along a filament and thus
over-emphasizes the collapse.
}\bigskip
%\label{mh.proj}
%\end{figure}

\end{document}